\documentclass[preprint,12pt]{elsarticle}
\usepackage{psfrag}
\usepackage[T1]{fontenc}
\usepackage{epsfig}
\usepackage{times}
\usepackage{amsmath}
\usepackage{amssymb}
\usepackage{graphicx}
\usepackage{wasysym}
\usepackage{textcomp}
\input epsf.sty
\newcommand{\be}{\begin{equation}}
\newcommand{\ee}{\end{equation}}
\newcommand{\ba}{\begin{eqnarray}}
\newcommand{\ea}{\end{eqnarray}}
\newcommand{\bi}{\begin{itemize}}
\newcommand{\ei}{\end{itemize}}

\renewcommand{\vec}[1]{{\bf #1}}

\def\lsi{\raise0.3ex\hbox{$<$\kern-0.75em\raise-1.1ex\hbox{$\sim$}}}
\def\gsi{\raise0.3ex\hbox{$>$\kern-0.75em\raise-1.1ex\hbox{$\sim$}}}

\newcommand{\BE}{\begin{eqnarray}}
\newcommand{\EE}{\end{eqnarray}}
\newcommand{\BDM}{\begin{displaymath}}
\newcommand{\EDM}{\end{displaymath}}

\makeatletter \@addtoreset{equation}{section} \makeatother

\begin{document}

\begin{frontmatter}

\title{Monte Carlo Search for Very Hard KSAT Realizations \\ for Use in Quantum Annealing}

\author[jsc]{T. Neuhaus}

\address[jsc]{J\"ulich Supercomputing Centre,
Forschungszentrum J\"ulich, D-52425 J\"ulich, Germany}
%\address[bie]{Fakultät für Physik, Universit\"at Bielefeld, D-33501 Bielefeld, Germany}

%\email{t.neuhaus@fz-juelich.de}

\begin{abstract}
%\input{abstract.tex}
% checked november
Using powerful Multicanonical Ensemble Monte Carlo methods from 
statistical physics we explore the realization space of 
random K satisfiability (KSAT) in search for computational 
hard problems, most likely the 'hardest problems'. 
We search for realizations with unique satisfying assignments 
(USA) at ratio of clause to spin number $\alpha=M/N$ that is 
minimal. USA realizations are found for $\alpha$-values that 
approach $\alpha=1$ from above with increasing number 
of spins $N$. We consider small spin numbers in $2 \le N \le 18$. 
The ensemble mean exhibits very special properties.
We find that the density of states of the first
excited state with energy one $\Omega_1=g(E=1)$
is consistent with an exponential divergence in $N$:
$\Omega_1 \propto {\rm exp} [+rN]$.
The rate constants for $K=2,3,4,5$ and $K=6$ of KSAT 
with USA realizations 
at $\alpha=1$ are determined numerically to be in the
interval $r=0.348$ at $K=2$ and $r=0.680$ at $K=6$. 
These approach the unstructured search value ${\rm ln}2$ with increasing $K$. 
Our ensemble of hard problems is expected to provide a test bed 
for studies of quantum searches with Hamiltonians that have 
the form of general Ising models.
\end{abstract}

%\pacs{75.10.Nr, 02.70.Ss, 03.67.Ac, 64.70.Tg}
\end{frontmatter}

\section{Introduction}

Random satisfiability problems like three satisfiability 
(3SAT) and its generalization KSAT form a corner stone of 
complexity theory, a very active research branch in formal 
logic and computer science. In these theories one is 
concerned with logical forms ${\cal F}(X)$ defined on some
bit space $\vec{X}$ and one discusses the question whether or not
there exists an assignment $\vec{X}_0$ that turns the value
of the logical form ${\cal F}(\vec{X}_0)$ into ``true''. The 
decision problem of KSAT and its accompanying function problem: the 
actual calculation of $\vec{X}_0$ at given ${\cal F}(\vec{X})$ 
for $K \ge 3$ belong to the class of NP complete theories \cite{Cook}, 
which for all practical purposes implies computational intractability. 
In these theories it is very common that worst case 
realization ensembles of forms ${\cal F}(\vec{X})$ exhibit an 
algorithm dependent complexity $C$, that rises 
exponentially $C \propto {\rm exp}[+rN]$ with the 
number of bits $N$. The rate constants $r$ are smaller
than the unstructured search value $r = {\rm ln} 2$ but at the
same time can take values that are finite fractions of ${\rm ln}2$. 
This implies, that there exist problems which are not solvable 
even for small numbers of bits like $N=100$, neither by 
analytic nor numeric methods, even usung brute computational force.

It is the privilege of statistical physics to turn the 
abstract notion of satisfiability into studies of  
Hamiltonian systems upon mapping the bit degrees of freedom $X_i=0,1$ 
via $s_i=2X_i-1$ for $i=1,...,N$ to Ising degrees of freedom 
$s_i \pm 1$, and upon introducing a suitable Hamiltonian $H_{\rm KSAT}$ 
whose ground-states at energy $E=0$ map one by one to the 
satisfying assignments of ${\cal F}(\vec{X})$. One may either 
consider classical statistical 
physics where the theory is supplied by artificial thermal 
fluctuations at inverse temperature $\beta=T^{-1}$ within the framework 
of the canonical partition function 
$Z_{\rm C}=\sum_{\rm Conf.} {\rm exp}[-\beta H_{\rm KSAT}]$ or 
alternatively, consider the quantum statistical theory of Pauli spins 
$S_i^x,S_i^y$ and $S_i^z$ with the quantum partition function 
\be
Z_{\rm Q}={\rm Tr}<\Psi  \mid {\rm exp}[-\beta [(1-\lambda)\sum_i S_i^x + \lambda H_{\rm KSAT}(S_i^z)]]\mid \Psi >, 
\label{quantum_partition}
\ee
where quantum fluctuations  at low $T \approx 0$ are tuned 
via an external parameter $\lambda$. For both 
cases the mathematical intractability is encoded into physical theories and 
it is an exciting research topic to study its 
consequences i.e., phase transitions and correlations
from various points of view. For the classical theory it was
shown, that computational intractability is related to
a phase transition - the SAT transition - along the principal parameter
direction $\alpha=M/N$ of random KSAT theories \cite{Kirkpatrick_1996}, the ratio
$\alpha$ hereby denoting the ratio of clause $M$ to spin $N$ numbers. 
In a later effort complexity related observables were determined
analytically within the framework of replica symmetry breaking theory
for random 3SAT \cite{parisi_science_2002}, and also numerically in large scale simulations \cite{mann_hartmann_02009}.
In particular the critical point of the 3SAT transition was determined to be
$\alpha_S=4.267...$ analytically. For the quantum theory, and within quantum information 
theory it was conjectured that adiabatic quantum computations (AQC) based 
on the properties of $Z_{\rm Q}$ could possibly obtain ground states of
$H_{\rm KSAT}$ in polynomial physical time \cite{Nishimori,Farhi}. For hard 3SAT realizations 
it turned out however, that early findings on polynomial 
ground state search times had to be corrected to exponentially 
large ones \cite{neuhaus_02011} for the simplest case of AQC making use of a
transverse magnetic field and a linear $\lambda$-parameter schedule. A similar 
finding was made recently for a another satisfiability theory: Exact Cover \cite{Young_both}.

\begin{figure}[t]
\centering
\includegraphics[angle=-90,width=10.0cm]{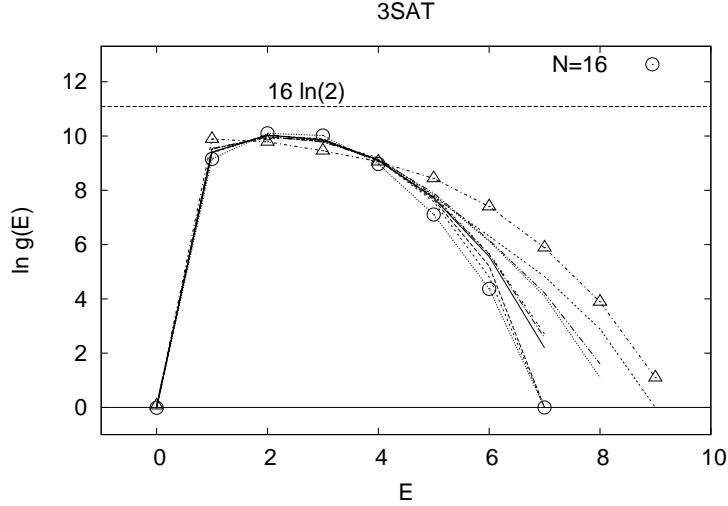}
%\caption{We display an example for Logarithmic scale density of states functions
%$g^\eta(E)$ for $\eta=1,...,10$ realizations within our hard problem realization
%set and for the theory 3SAT at $N=16$ spins. Density of states functions
%have finite support at discrete energy values $E=0$ (ground-state)
%and $E=1,2,3...$ but for optical reconnaissance reasons are connected with polygons.
%Some functional values are identified by circles and triangles.
%The total number of spins configurations at $N=16$ is $2^{16}=65536$. It is remarkable: 
%the density of states jumps from $g(E=0)=1$ (USA) to a large value 
%$\Omega_1=g(E=1) \approx e^{9.3}\approx 11000$. Within the paper we find clear numerical evidence
%for an exponential singularity in $\Omega_1$ as a function of $N$:  
%$\Omega_1 \propto {\rm exp}[+rN]$ with $r=0.567$ in 3SAT. 
%As far as ground-state searches are concerned: any stochastic ground-state search can
%easily reach the $E=1$ surface. Beyond that and in front of $E=0$ the search has to
%enumerate an exponential large number of possibilities. 
%\label{figure_1}}
% SHORTENED THE THING
\caption{We display an example for Logarithmic scale density of states functions
$g^\eta(E)$ for $\eta=1,...,10$ realizations for the theory 3SAT at $N=16$ spins. 
Density of states functions
have finite support integer values but for optical 
reconnaissance reasons are connected with polygons.
Some functional values are identified by circles and triangles.
%The total number of spins configurations at $N=16$ is $2^{16}=65536$. 
It is remarkable: the density of states jumps from $g(E=0)=1$ (USA) to a large value 
$\Omega_1=g(E=1) \approx e^{9.3}\approx 11000$. 
%Within the paper we find clear numerical evidence
%for an exponential singularity in $\Omega_1$ as a function of $N$:  
%$\Omega_1 \propto {\rm exp}[+rN]$ with $r=0.567$ in 3SAT. 
As far as ground-state searches are concerned: any stochastic ground-state search can
easily reach the $E=1$ surface. Beyond that and in front of $E=0$ the search has to
enumerate an exponential large number of possibilities. 
\label{figure_1}}
\end{figure}

Within the current work we execute a very use-full exercise prior to the actual 
studies of complexity related observables in physical theories. We restrict 
the admissible set of all KSAT Hamiltonians, namely random KSAT realizations
with ensemble mean \break $<...>_{\rm RANDOM~KSAT}$, to a 
much smaller 'hard' set $<...>_{\rm HARD}$
of $H_{\rm KSAT}^\eta$ Hamiltonian's with corresponding ensemble mean. 
The index $\eta$ denotes the ensemble members which for reasons of 
computer time limitations have finite number $\eta=1,...,1000$ throughout the paper. 
As far as ground-state searches are concerned our problem set is targeted
at hard problems - most likely the 'hardest problems' - which otherwise and 
within $<...>_{\rm RANDOM~KSAT}$ are exponentially rare.
Our problems are constructed under specific constraints:  
\begin{itemize}
\item  The ground-state to any $H_{\rm KSAT}^\eta$ is unique, which if $g^\eta(E)$ denotes the
density of states function (DOS) implies $g^\eta(E=0)=1$. Such problem realizations possess unique
satisfying assignment's (USA).
\item For a given number of spins $N$ and for realizations with 
$g(E=0)=1$ the number of clauses $M$ is minimal.
The parameter $\alpha$ is then minimal too $\alpha=\alpha_{min}$. E.g. : we find that USA realizations
in 3SAT for $\alpha_{min}$ follow $\alpha_{min}=(N+4)/N$.
\item The set of problem realizations  $<...>_{\rm HARD}$ is drawn with
unique probability from the set of $<...>_{\rm RANDOM~KSAT}$ realizations.
\end{itemize}
Similar realizations have lately been considered for 3SAT in Ref. \cite{Znidaric_2005} 
with a weaker constraint on the value $\alpha$, which had the value  
$\alpha=3$. The minimal KSAT values within this work turn out to approach $\alpha_{min}=1$ from above 
independent of $K$ with increasing $N$. In short: we are constructing USA realizations
in KSAT at $\alpha=1$ asymptotically.

%At the heart of our numerical calculations is a Markov Chain Monte Carlo
%study of the ground state multiplicties $\mu$ generating constraint partition function
%\be
%\Gamma(\mu)= {\cal N}^{-1}  \sum_{\rm Random~3SAT} \delta^{(1)}[\mu - g(E=0) ],
%\ee
%that utilizes flat histogram sampling methods i.e., we use a biased Monte Carlo scheme
%which lifts probability depreciated $\mu$ sectors within $\Gamma$.
At the heart of our numerical calculations is a Markov Chain Monte Carlo
study of the partition function
\be
\Gamma(\mu)= {\cal N}^{-1}  \sum_{\rm Random~KSAT} \delta^{(1)}[\mu - g(E=0) ],
\ee
which partitions the realization space of random KSAT
within respect to $\mu$, the ground-state multiplicity. 
Once the Markov Chain Monte Carlo visits the $\mu=1$ sector (USA)
corresponding problems are collected on the disk of a computer. Similar
flat histogram sampling methods, like Wang-Landau \cite{WangLandau} and Multicanonical \cite{Muca} 
simulations have recently been used in complexity theory in an attempt to sample
the density of states function $g(E)$ in 3SAT for spin numbers $N$ that prohibit
exact enumeration \cite{random}. The final part of the paper classifies measures of complexity
within typical problem realizations in $<...>_{HARD}$.

Today's understanding on the origin on the complexity 
of the physical search in frustrated and disordered systems pictures 
a free energy landscape in which as a function of the value $N$, a finite number 
of solution clusters is accompanied by an exponentially large number of almost 
solution clusters at energy near but above the ground-state.  All clusters 
are separated by finite free energy barriers. The situation resembles 
the search for a needle - or several needles - in a haystack.  
%In particular one expects, that a classical search, which
%probes the landscape by local spin moves, needs an exponential large 
%number of trials to actually find the ground-state, or its vicinity. 
A simplified mechanism operates within 
our hard problem ensemble $<...>_{\rm HARD}$. 
%We find, that density of states functions $g^\eta(E)$ 
%for typical $\eta$ realizations exhibit an almost discontinuous i.e., step-like 
%behavior if the energy increased from the ground state by just one unit - the energy gap - 
%to energy $E=1$ i.e., our theory tends to accumulate spin-configurations at the energy value one, while
%$g(E=0)=1$ is kept finite at unity. An example is displayed in Fig.(\ref{figure_1}). 
We find, that the phase space volume $\Omega_1$ at $E=1$
is exponentially large in the number of degrees of freedom $N$, see the examples of $g(E)$ 
displayed in Fig.(\ref{figure_1}). Thus first: for all 
of the considered KSAT theories with $K=2$ up to $K=6$ we encounter 
the generic situation: a single needle is searched in a haystack of exponential large 
size \footnote{The theories at $K \ge 3$ are NP-complete
while at $K=2$ there exist mathematical polynomial time algorithms that find the ground-state even though $\Omega_1$ is exponentially large.}. Second: we find numeric evidence that actual 
values of $\Omega_1$ are extremal i.e., maximal
under the condition of minimal $\alpha$, which in turn justifies the notion of
most likely the 'hardest problems'.

%\newpage

\section{Theory, Hard Problems and Monte Carlo Simulation}

\subsection{Theory and Observables}
 
 In KSAT one considers logical forms ${\cal F}$ - a 
function - whose truth value can either be true or false 
and which are defined on a 
space of $N$ Boolean degrees of freedom - bits - $X_i$ with $i=1,...,N$. 
In the satisfiability problem one asks for the existences of assignment's 
i.e., bits $\vec{X}_0$ that would evaluate the function ${\cal F}$ at 
the value true.
Solving the function problem implies the explicit calculation of a
single satisfying assignment or, of all different satisfying assignments if 
there are several of those. The logical form ${\cal F}$ is the conjunctive normal
form of $M$ clauses $\lbrace C_1,...,C_M \rbrace$: 
${\cal F}=C_1 \land C_2 \land ... \land C_M$, which only evaluates
true if all clauses $C_\alpha$ with $\alpha=1,...,M$ evaluate true
simultaneously. Any of the $M$ clauses is the disjunction of integer 
$K$ literals  $L_{\alpha,j}$ with $K \ge 2$ and $j=1,...,K$:
\begin{equation}
C_\alpha= L_{\alpha,1}  \lor  L_{\alpha,2} \lor ... \lor L_{\alpha,K}.
\end{equation} 
A clause is true, if at least one of its literals 
evaluates true. For example, in 3SAT there are $7$ configurations of 
literals on the clause which evaluate true and just one with 
truth value false. In addition, a literal is either a bit $X$ 
or its negation $\overline{X}$ and, the actual identification
of a literal with a specific bit - or its negation - is controlled
by a map $(\alpha,j) \rightarrow i:i=i[\alpha,j]$, that associates 
clauses and clause-positions $\alpha,j$ to the index set $i$ of 
bits. The map $i=i[\alpha,j]$ and the possibility
of $2^{KM}$ negations at the literal positions 
are free parameters of the theory. In an
Hamiltonian theory they can be used to introduce ensembles with mean $<...>$
over random disorder as well as random frustration, a possibility that is 
heavily exploited in this work. It is implicitly understood, that
tautologies i.e., contradicting pairs within clauses 
like $X_i\overline{X}_i$ as well as redundancies i.e., duplicate literals like 
$X_i X_i$ or  $\overline{X}_i \overline{X}_i$ are not admitted to 
the theory.

\begin{figure}[t]
\centering
\includegraphics[angle=-90,width=10.0cm]{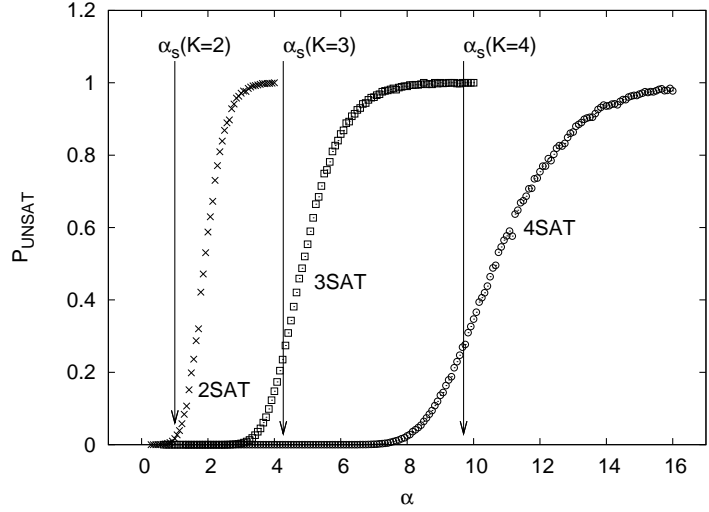}
\caption{Probability $P_{\rm UNSAT}$ of un-satisfiable formulas 
within  $<...>_{\rm RANDOM~KSAT}$ for $K=2,3$ and $K=4$ als a function of $\alpha$. Exact and numerical values
for the SAT to UNSAT threshold $\alpha_s(K)$ are indicated by arrows. The numerical data are obtained
from the partition function $\Gamma(\mu)$ of eq.(\ref{partition_function}) via eq.(\ref{eq:unsat}).
\label{fig:punsat}}
\end{figure}

 The physical degrees of freedom are classical Ising spins $s_i= \pm 1$
with $i=1,...,N$ and without loss of generality, true on each
bit $X_i$ is identified with spin up $s_i=+1$.
Let us introduce functions $h_\alpha$ in an attempt
to write the Hamiltonian $H_{\rm KSAT}$ as a sum of $M$ terms:
$H_{\rm KSAT}=\sum_{\alpha} h_\alpha$, where each term 
corresponds to a clause and, where the ground-states 
of $H_{\rm KSAT}$ at energy $E=0$ can be identified one by one
with the satisfying assignments of ${\cal F}$. For this purpose we note that 
$K$ spins $s_1,...,s_K$ of the clause $C=X_1  \lor  X_2 \lor ... \lor X_K$ 
add up to the sum $\Sigma=\sum_{i=1}^K s_i$, which takes 
$K+1$ different values 
$\Sigma =-K,-K+2,...,K-2,K$. Consequently the 
polynomial $h=h(s_1,...,s_K)$
\begin{equation}
h={ (-1)^K \over 2 ^K K!}\prod_{m=1}^K( \sum_{i=1}^N s_i +K-2m)
\end{equation}
has the value $h=0$ for all spin configurations except one,  
if only all spins are down: $s_i=-1$ with $i=1,...,N$. For the latter case $h=1$, which 
implies an energy-gap of value unity. For $K \ge 2$ the function
$h$ is a linear combination of the spins n-point functions
$\Gamma^0$, $\Gamma^1$, ... , $\Gamma^K$ with a maximum $n$ 
of value $n_{max}=K$. For purposes of illustration 
we present the 2SAT and 
3SAT cases. For 2SAT we obtain the anti-ferromagnet at finite field
\begin{equation}
h_{\rm 2SAT}={1 \over 4}[s_1 s_2 - (s_1 + s_2) +1],
\end{equation}
while in 3SAT
\begin{equation}
h_{\rm 3SAT}={1 \over 8}[ s_1 s_2 s_3 + (s_1 s_2 + s_1 s_3 + s_2 s_3) 
+ (s_1 + s_2 + s_3) - 1 ].
\end{equation}
The necessary frustrations are encoded in a 
matrix array $\epsilon_{\alpha,j} = \pm 1$ which for each clause 
$\alpha$ and position $j$ with $j=1,...,K$ follows the pattern of 
negations within ${\cal F}$, a negation induces an $\epsilon=-1$ while otherwise
$\epsilon=+1$. We mention that in random KSAT, which we denote by 
the ensemble mean  $<...>_{\rm RANDOM~KSAT}$, values of $\epsilon$ are drawn with
equal probability $p(\epsilon=+1)=p(\epsilon=-1)={1 \over 2}$. The final
form of the KSAT Ising Hamiltonian $H_{\rm KSAT}$ is
\begin{equation}
H_{\rm KSAT} =  \sum_{\alpha=1}^M h_{\rm KSAT}
( \epsilon_{\alpha,1}s_{i[\alpha,1]} , \epsilon_{\alpha,2}s_{i[\alpha,2]} , ... , 
\epsilon_{\alpha,K-1}s_{i[\alpha,K-1]} , \epsilon_{\alpha,K}s_{i[\alpha,K]} ),
\label{the_hamiltonian}
\end{equation}
and is the basis of our studies. Its principal parameters for $K$ are the ratio 
of clause numbers $M$ over $N$ namely $\alpha=M/N$, and the particular assignments
of spins to clauses via the map $i[\alpha,j]$, as well as the settings within the 
frustration matrix $\epsilon_{\alpha,j} = \pm 1$. We denote a specific setting
of the latter map and matrix a realization and study ensemble mean 
expectation values of observables at fixed $\alpha$ throughout the paper.

\begin{figure}[t]
\centering
\includegraphics[angle=-90,width=10.0cm]{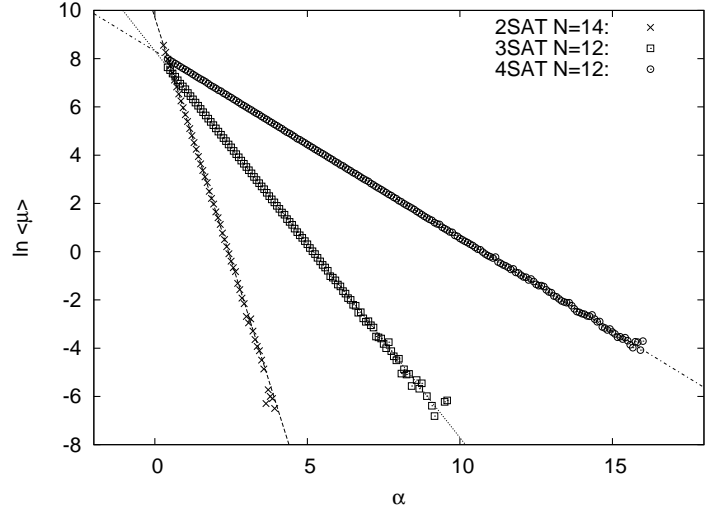}
\caption{Monte Carlo data for ${\rm ln}<\mu>$ of random KSAT
in accord with eq.(\ref{eq:mean_mu}) for selected $N$ values 
as a function of $\alpha$ and for $K=2,3$ and $K=4$.
The straight lines match the Monte Carlo data and
correspond to the exact result of eq.(\ref{eq:mean_mu_exact}).
\label{fig:mean_mu}}
\end{figure}

Once the Hamiltonian is given we formally define the
canonical partition function $Z(\beta)=\sum_{\rm Conf.}e^{-\beta H}$
which at temperature $T=\beta^{-1}$ allows the definition of physical observables
as there are the internal energy $U=\partial_\beta {\rm ln} Z$, or the specific heat  
$C_V=\beta^{2}\partial_\beta U$. The canonical partition function has the
spectral representation
\begin{equation}
Z(\beta) = \sum_E g(E) e^{-\beta E},
\label{eq:canonical}
\end{equation}
where $g(E)$ denotes the density of states (DOS):
\begin{equation}
g(E)=\sum_{\rm Conf.}\delta^{(1)}(H-E).
\label{eq:dos}
\end{equation}
For KSAT theories $g(E)$ is integer valued, has finite support
on the integer values of the compact interval $0 \le E \le M$ and an integral
$\sum_E g(E)=2^N$. A satisfiable Boolean form induces $g(E=0)>0$, while 
$g(E=0)=1$ corresponds 
to an ${\cal F}$ that only has one unique satisfying 
assignment (USA). Boolean forms, that cannot be satisfied have $g(E=0)=0$.
The quantity $g(E=1)$ also is denoted the microcanonic phase space volume $\Omega_1$
of the energy one energy surface.

\begin{figure}[t]
\centering
\includegraphics[angle=-90,width=10.0cm]{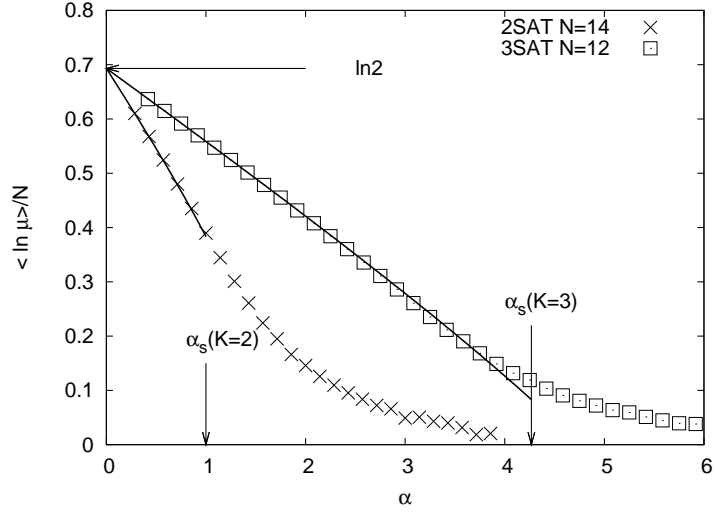}
\caption{Ground state entropy density $s_0={1 \over N}<{\rm ln } {\mu} >|_{\mu >0}$
for random KSAT at $K=2$ and $K=3$ as a function of $\alpha$.
The arrows denote exact positions of the SAT to UNSAT
threshold at $\alpha_s$. The data sets are superimposed by 
series expansion results for $s_0$ for $\alpha$-values below $\alpha_s$. The curves  
lie on top of the data.
\label{fig:s0}}
\end{figure}

Our knowledge of the statistical properties of K satisfiablity stems 
from extensive analytical \cite{parisi_science_2002} and numerical studies \cite{Kirkpatrick_1996}
of random KSAT, which 
have demonstrated the existence of a transition, possibly a phase transition
at values $\alpha_s(K)$. The SAT to UNSAT transition separates at low 
$\alpha<\alpha_s$ a phase where formulas ${\cal F}$ are satisfied 
in the mean, from a phase at large $\alpha>\alpha_s$ where formulas ${\cal F}$ can 
not be satisfied. Numerical data for the probability 
$0 \le P_{\rm UNSAT} \le 1$ of un-satisfiable formulas within the mean of random KSAT 
are displayed in Fig.(\ref{fig:punsat}) and illustrate the statement.
The data are of similar quality as the data obtained by Selman and Kickpatrick in 1996 \cite{Kirkpatrick_1996}.
The consensus is that probable realizations within random KSAT are 'hardest', i.e. 
computational most intractable, at and in the vicinity of the transition 
point $\alpha \approx \alpha_s$. However, this does not 
exclude the existence of still 'harder' i.e., worst case realizations which at arbitrary
$\alpha$ are hidden in the tails of probability distribution functions for complexity related observables
with small, possibly very small probabilities.

\subsection{Search for Hard Problems}

The starting point of our search for 'hard' realizations are
observations that concern realizations with USA. If one considers USA realizations
in 3SAT for the smallest spin number $N=3$ and clause number $M$ 
one inevitably arrives at the $M=7=3+4$ realization
\begin{equation}
 \begin{array}{cccc}
		\mathcal F_{\rm USA} (N=3) = 	&(&1 &\lor \quad 2 \quad \lor \quad 3 \quad) \quad \land \\
				&(&1 &\lor \quad 2 \quad \lor \quad \overline{3} \quad) \quad \land \\
				&(&1 &\lor \quad \overline{2} \quad \lor \quad 3 \quad) \quad \land \\
				&(&1 &\lor \quad \overline{2} \quad \lor \quad \overline{3} \quad) \quad \land \\
				&(&\overline{1} &\lor \quad 2 \quad \lor \quad 3 \quad) \quad \land \\
				&(&\overline{1} &\lor \quad 2 \quad \lor \quad \overline{3} \quad) \quad \land \\
				&(&\overline{1} &\lor \quad \overline{2} \quad \lor \quad 3 \quad) \quad ,  ~~~\\
\end{array}
\label{eqn:f3_clause}
\end{equation}
which encodes the unique ground state $s_1=s_2=s_3=+1$. This particular example is one of 
eight that all encode USA's for $N=3$, and is turned in a readable 
form upon permuting clause and literal indices. It has 
interesting specific properties:
\begin{itemize}

\item{} FOR $N=3$ ${\cal F}_{\rm USA}$ is the minimal form with
a USA. For $N=3$ and $M=6$ there are no USA realizations in 3SAT.

\item{} The density of states $g(E)$ only has two values
$g(E=0)=1$ and $g(E=1)=7$. All spin flips acting on the ground-state lift the $E=0$ energy surface 
by just one unit to $E=1$. The states with $E=1$ have dis-proportional 
large multiplicity and therefore $E=0$ is hidden.
This suggests that still 'minimal' but larger forms ${\cal F}_{\rm USA}$ at values $N>3$ 
could inherit a similar property. These must exist at $\alpha= (N+4)/N$
as one can introduce additional spins and clauses one by one. For example, if
we introduce a fourth spin and extend ${\cal F}_{\rm USA}$  
by one clause to an $(N,M)=(4,8)$ form with comparable property, then
\begin{equation}
 \mathcal F_{\rm USA}(N=4) \quad =  \quad \mathcal F_{\rm USA}(N=3) \quad \land \quad (\quad 4 \quad \lor \quad \overline{1} \quad \lor \quad \overline{2} \quad). 
\label{eqn:f3_clause_two}
\end{equation}  
The latter form encodes the unique ground state  $s_1=s_2=s_3=s_4=+1$ and 
has the density of states $g(E=0)=1$ and $g(E=1)=15$ respectively. Again $E=1$ configurations have 
large multiplicity. 

\item{} Within ${\cal F}_{\rm USA}$ of eq.(\ref{eqn:f3_clause}) there are exactly 
$m_{1}=3$ clauses - those with two negations - which in the unique 
solution are solved by just one true literal. There are in addition $m_2=3$ clauses which are solved by
two literals and $m_3=1$ clauses which are solved by three literals. 
Also, there exists a polynomial transformation of 3SAT to maximal independent 
set (MIS) \cite{mis}. It is easy to show, that a unique ground-state of the 3SAT problem
transforms into a degenerate ground-state  
in the corresponding MIS problem. The ground-state multiplicity MIS, $\Omega_{0,{\rm MIS}}$, has the value
\begin{equation}
\Omega_{0,{\rm MIS}}=2^{m_2}3^{m_3}=24 ~~~~ (3SAT,N=3,M=7),
\end{equation}
on  ${\cal F}_{\rm USA}$.
We note that $m_1$ of ${\cal F}_{\rm USA}(N=4)$ 
turns out to be $m_1=4$ while $m_2$ and $m_3$ remain having values $m_2=3$ and $m_3=1$,  
and thus also ${\cal F}_{\rm USA}(N=4)$ is constant at $\Omega_{0,{\rm MIS}}=24$. It is suggested that
'minimal' but larger ($N>3$) forms ${\cal F}_{\rm USA}$ can have indices $m_1$
that are of magnitude ${\cal O}(N)$, which in turn limits the volume $\Omega_{0,{\rm MIS}}$
to finite values. Finite values imply vanishing
ground state entropy density ${\rm ln}\Omega_{0,{\rm MIS}} /N$ under the polynomial transformation from 3SAT to MIS.

\end{itemize}
\begin{figure}[t]
\centering
\includegraphics[angle=-90,width=10.0cm]{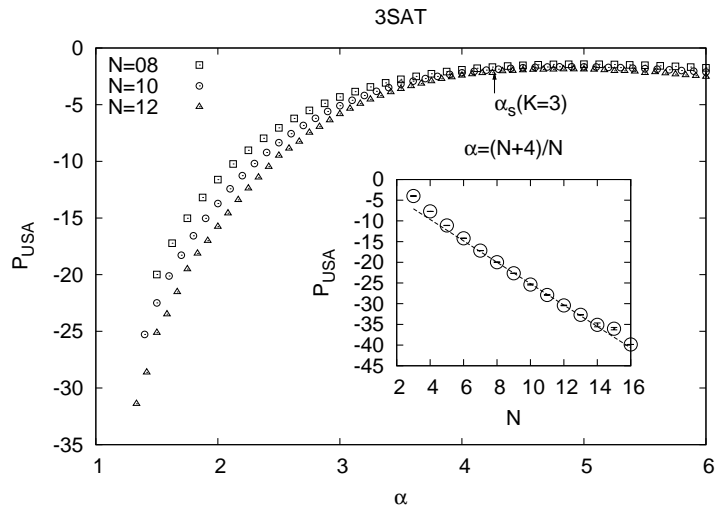}
\caption{ Probability $P_{USA}$ of eq.(\ref{eq:pusa}) for the occurrence
of problem realizations with unique satisfying assignment (USA) in 3SAT as a function
of $\alpha=M/N$. Spin numbers are $N=8,10$ and $N=12$. The inset of the figure
displays the decrease of $P_{USA}$ for $\alpha_{\rm HARD}$, see eq.(\ref{alpha_minimum})
as a function of $N$.
\label{fig:pusa}}
\end{figure}
The existence of examples with interesting properties guides our expectations.
The question is raised whether USA and 
3SAT realizations at the ratio of clause to spin numbers 
\begin{equation}
\alpha_{\rm HARD}= { N+4 \over N}
\end{equation}
exist for arbitrary $N \ge 3$ and what their properties are ? In absence of
useful mathematical methods 
%for the evaluation of the partition function of eq.(\ref{partition_function})
we use Monte Carlo simulations in order to actually construct members of 
the ensemble at $\alpha_{\rm HARD}$, and in a later measurement step we determine their properties. In
particular we calculate $\Omega_1$, the multiplicity of the energy one surface. It is then
necessary to employ biased Monte Carlo sampling techniques, as in the vicinity of $\alpha=1$
USA realizations within random 3SAT have exponentially small probability. Finally it 
is easy to generalize our arguments to arbitrary $K$. For KSAT 
with $K \ge 2$ we expect USA realizations with minimum clause number at
\begin{equation}
\alpha_{\rm HARD}(K)= { N + 2^K - K - 1   \over N},
\label{alpha_minimum}
\end{equation}
under the condition that $N \ge K$.

\subsection{Monte Carlo Search and Checks}

The Monte Carlo simulation performs a stochastic estimate of the biased partition function 
\begin{equation}
\Gamma(\mu,W_{\rm MUCA})={\cal N}^{-1}\sum_{\rm RANDOM~KSAT} e^{+W_{\rm MUCA}(\mu)}
\delta^{(1)}[\mu -g(E=0)],
\label{biassed_partition_function} 
\end{equation}
which for $0 \le \mu \le 2^N$ is evaluated on the phase space 
of all possible random KSAT realizations 
for a given KSAT Hamiltonian eq.(\ref{the_hamiltonian}). The bias, as expressed
by the Boltzmann factor ${\rm exp}[+W_{\rm MUCA}(\mu)]$, is introduced along the lines of Multicanonical
Ensemble simulations \cite{Muca} and serves the purpose to lift the probabilities of rare $\mu$
configurations in the Markov chain. The Monte Carlo is expected to perform a random walk in $\mu$
and whenever the $\mu=1$ sector is visited an ensemble member of $<...>_{\rm HARD}$ 
is stored on the disk of a computer. Our Monte Carlo is quite un-conventional
and essential remarks are in order:

\begin{itemize}

\item{} The Markov chain of configurations consists of realizations as specified
by their maps $i[\alpha,j]$ and frustration matrix $\epsilon_{\alpha,j}$. Each
problem realization is attached to a Hamiltonian
theory with density of states $g(E)$ that can be evaluated at $E=0$,  $\mu=g(E=0)$. The calculation
of $\mu$ for a given configuration unfortunately takes ${\cal O}(2^N)$ computational steps. Our Monte Carlo
simulation therefore is limited to small numbers of spins. We studied KSAT theories for $K=2,3,4,5$ and $K=6$.
We were able to generate ensembles of $1000$ statistical independent members each for maximum
spin numbers $N_{\rm max}=18,16,14,12$ and $N_{\rm max}=10$ respectively. Minimum spin numbers
always are $N_{\rm min}=K$..

\item{} Configurations are updated with Metropolis updates \cite{Metropolis_Rosenbluth}. The 
initial problem realization at $\mu_{\rm I}$ is subject to a trial-update 
which targets $\mu_{\rm F}$. The Markov chain accept probability for the move is
\begin{equation}
P_{\rm ACC} = {\rm min}[1,e^{+W_{\rm MUCA}(\mu_{\rm F})-W_{\rm MUCA}(\mu_{\rm I}) }],
\end{equation}
and as usual, if the update is rejected the initial configuration stays within the Markov Chain.

\item{} Trial updates are generated randomly on the space of random 
KSAT realizations. One chooses a random clause $\alpha_0$ and clause position $j_0$ 
and at $(\alpha_0,j_0)$ trial values $i_{\rm Trial}$ and $\epsilon_{\rm Trial}$, which are
uniformly distributed on the measure of the theory. The absence of redundancies and tautologies constrains the
admissible move set. The typical number of Monte Carlo moves for the simulation
of $\Gamma(\mu)$ is $10^9$. For the larger $N$ values it was necessary to repeat 
the simulations with different random number sequences possibly 10, up to several 10 times. 
The numerical data, as presented in the paper, consumed
one month of computer time on a $256$ processor workstation cluster. 

\item{} The bias $W_{\rm MUCA}(\mu)$ has to be chosen properly in order to guarantee
efficient random walk behavior in the variable $\mu$. In a preparation step we use 
Wang Landau simulations \cite{WangLandau} to generate sufficiently accurate $W_{\rm MUCA}(\mu)$ 
weight functions, which then enter the 
Multicanonical simulation of eq(\ref{biassed_partition_function}). 

\end{itemize}
The biased partition function of eq.(\ref{biassed_partition_function}) serves 
as a tool to facilitate Monte Carlo sampling of different $\mu$ sectors in random KSAT and  
in particular the sector $\mu=1$ (USA) is sampled efficiently. There is however an additional
benefit.  After finishing the biased Monte Carlo simulation a final reweighing 
step $\Gamma(W=0)={\rm exp}[-W(\mu)]\Gamma(W(\mu))$ restores the un-biased partition function
\begin{equation}
\Gamma(\mu)={\cal N}^{-1}\sum_{\rm RANDOM~KSAT} \delta^{(1)}[\mu -g(E=0)], 
\label{partition_function}
\end{equation}
which on the space of random KSAT realizations simply counts the probability
of $E=0$ multiplicities $\mu$. Given $\Gamma(\mu)$ we can determine expectation 
values of known observables within random KSAT, which provide consistency checks 
on the correctness of the Monte Carlo simulation. A list and a comparison
to numerical data follows:

\begin{itemize}

\item{} In random KSAT there is always a finite
probability of problem realizations with $E=0$ non-vanishing multiplicity. 
In fact one can calculate the $E=0$ mean
multiplicity 
\begin{equation}
<g(E=0)>_{\rm RANDOM~KSAT} =  <\mu> ={\cal N}^{-1} \sum_{\mu} \Gamma(\mu)\mu  
\label{eq:mean_mu}
\end{equation}
on combinatorial grounds at arbitrary $K$ exactly \cite{random}, which simply yields
\begin{equation}
<\mu> =(1-{1 \over 2^K})^M 2^N .
\label{eq:mean_mu_exact}
\end{equation}
In Fig.(\ref{fig:mean_mu}) we compare selected measurement data for $<\mu>$
with the exact result for various values of $K$, $N$ and $M$. The Monte Carlo data agree 
with the combinatorial result very well.

\item{} One may wonder whether a theory with an entirely regular $<\mu>$ 
will contain a non-regular structure at the SAT to UNSAT transition $\alpha_s$. 
However, the constraint expectation value of the quantity 
${1 \over N}<{\rm ln } {\mu} >|_{\mu >0}$, under omission of the $\mu=0$ sector does 
in fact show non-trivial behavior.  Within the SAT phase 
$(\alpha < \alpha_s)$ it equals the ground-state entropy density
$s_0={1 \over N}<{\rm ln } {\mu } >$, for which long time ago \cite{Monasson_1997} and for the
theories 2SAT and 3SAT an $\alpha$ series-expansion was calculated 
within replica symmetry breaking theory up to order ${\cal O}(\alpha^{10})$. In Fig.(\ref{fig:s0}) we compare
our numerical data $<s_0>$ in 2SAT and 3SAT with the series expansions results. The figure
contains two curves, which for 2SAT for $\alpha <\alpha_s=1$ and for 3SAT for $\alpha <\alpha_s=4.267$
are indistinguishable from the numerical data points. Finally we note for random KSAT, that the probability
$P_{\rm UNSAT}$ of an un-satisfiable formula has the simple
representation 
\begin{equation}
P_{\rm UNSAT}=  {\cal N}^{-1}\Gamma(\mu=0).
\label{eq:unsat}
\end{equation}
The data are displayed in Fig.(\ref{fig:punsat}).
\end{itemize}
The main reason for the use of quite elaborate Monte Carlo techniques 
is the rareness of USA realizations for $\alpha \approx 1$, in particular
for the conjectured exact point $\alpha=\alpha_{\rm HARD}(K)$, as given in eq.(\ref{alpha_minimum}).
For all our theories with $K=2,3,4,5$ and $K=6$ and for typical $N$ like $N=10$
we search the $\alpha$-parameter space also at $\alpha$-values below $\alpha_{\rm HARD}(K)$ for USA
realizations. Neither Multicanonical Ensemble simulations for several weight functions $W_{\rm MUCA}$, nor Wang Landau
simulations or, alternatively simulated annealing runs in $\mu$ - ever 
produced a USA realization for $\alpha$ below $\alpha_{\rm HARD}$. 
However, at $\alpha_{\rm HARD}$ eq.(\ref{alpha_minimum}) USA realizations are found. 
The relative probability $P_{\rm USA}$ for the occurrence of 
unique satisfying assignment's within random KSAT is
\begin{equation}
P_{\rm USA}=  {\cal N}^{-1}\Gamma(\mu=1).
\label{eq:pusa}
\end{equation}
%We display in Fig.(\ref{fig:pusa}) $P_{\rm USA}$ data in 3SAT at $N=8,10,12$ spins and, all curves 
%terminate to the left at values $\alpha$ exactly equal to $\alpha_{\rm HARD}$. The probability
We display in Fig.(\ref{fig:pusa}) $P_{\rm USA}$ data in 3SAT for $N=8,10,12$ spins. 
$P_{\rm USA}$ appears to be a slowly varying function above $\alpha_s=4.267$, with a maximum around $\alpha_s$
and with a rapid decrease towards minimal and very small values at $\alpha_{\rm HARD}$ and, problems
with larger $N$ appear to be increasingly improbable. The asymptotic
decay of $P_{\rm USA}(\alpha_{\rm HARD})$ is consistent with an exponential decay
$P_{\rm USA}(\alpha_{\rm HARD}) \propto {\rm exp}(-rN)$ with $r \approx 2.58$ in 3SAT and is depicted
in the inset of Fig.(\ref{fig:pusa}). In addition at fixed spin number $N$
values of $P_{\rm USA}(\alpha_{\rm HARD})$ turn out to be even  smaller if larger 
$K$ values are considered. We quote ${\rm ln}P_{\rm USA}(\alpha_{\rm HARD})=-11.4,-31.39$ and ${\rm ln}P_{\rm USA}(\alpha_{\rm HARD}) \approx -84.3$
for the twelve spin theory with $K=2,3$ and in 4SAT. Finally we present 
for purposes of illustration a specific 3SAT realization for $N=16$ spins and $M=20$ clauses:
\begin{equation}
\begin{array}{cccccccccccccccccc}
    \mathcal F_{\rm USA} =  &(&  \overline{11}  &\lor& \quad            12   \quad&\lor&\quad   \overline{3}  &)& \quad \land &(&  \overline{14}  &\lor& \quad  \overline{13}  \quad&\lor&\quad   \overline{8}  &)& \quad \land \\  
                         &(&            11   &\lor& \quad             2   \quad&\lor&\quad            12   &)& \quad \land &(&   \overline{4}  &\lor& \quad             6   \quad&\lor&\quad            12   &)& \quad \land \\  
                         &(&             6   &\lor& \quad            12   \quad&\lor&\quad            13   &)& \quad \land &(&             6   &\lor& \quad            14   \quad&\lor&\quad   \overline{7}  &)& \quad \land \\  
                         &(&   \overline{8}  &\lor& \quad             6   \quad&\lor&\quad             9   &)& \quad \land &(&   \overline{5}  &\lor& \quad  \overline{12}  \quad&\lor&\quad             3   &)& \quad \land \\  
                         &(&  \overline{13}  &\lor& \quad  \overline{16}  \quad&\lor&\quad             4   &)& \quad \land &(&             8   &\lor& \quad             6   \quad&\lor&\quad            12   &)& \quad \land \\  
                         &(&             3   &\lor& \quad            12   \quad&\lor&\quad   \overline{6}  &)& \quad \land &(&             5   &\lor& \quad             3   \quad&\lor&\quad  \overline{12}  &)& \quad \land \\  
                         &(&            15   &\lor& \quad             5   \quad&\lor&\quad            12   &)& \quad \land &(&            12   &\lor& \quad            11   \quad&\lor&\quad   \overline{3}  &)& \quad \land \\  
                         &(&             6   &\lor& \quad  \overline{11}  \quad&\lor&\quad            15   &)& \quad \land &(&  \overline{15}  &\lor& \quad   \overline{3}  \quad&\lor&\quad  \overline{12}  &)& \quad \land \\  
                         &(&  \overline{13}  &\lor& \quad  \overline{15}  \quad&\lor&\quad            12   &)& \quad \land &(&            15   &\lor& \quad            16   \quad&\lor&\quad            10   &)& \quad \land \\  
                         &(&            15   &\lor& \quad   \overline{3}  \quad&\lor&\quad  \overline{12}  &)& \quad \land &(&             1   &\lor& \quad            12   \quad&\lor&\quad   \overline{9}  &)& \quad   ~~~ \\  
\end{array}
\label{eqn:3sat_clause_example}
\end{equation}
For $\alpha_{\rm HARD}=24/16=1.5$ it encodes the unique ground 
state $1100100111001000$ - zero corresponding to spin down and one corresponding to spin up - and
is characterized by the phase space volumes $\Omega_0=1$, $\Omega_1=19687$ and $\Omega_{0,{\rm MIS}}=24$. USA realizations
for the given parameter values have probability $P_{\rm USA} \approx 0.000000000000000004$ to occur by 
chance within random 3SAT. The full density of states of eq.(\ref{eqn:3sat_clause_example}) is 
depicted in Fig.(\ref{figure_1}), see the triangles in the figure. Finally the stochastic nature of the Monte Carlo search
result is apparent if one compares the random structure of eq.(\ref{eqn:3sat_clause_example})
with the regular structure in eq.(\ref{eqn:f3_clause}).

%\newpage

% \section{Results of Simulation}
\section{Properties of Hard KSAT Realizations}
\begin{figure}[t]
  \begin{minipage}[t]{0.5\textwidth}
    \includegraphics[angle=-90,width=8.0cm]{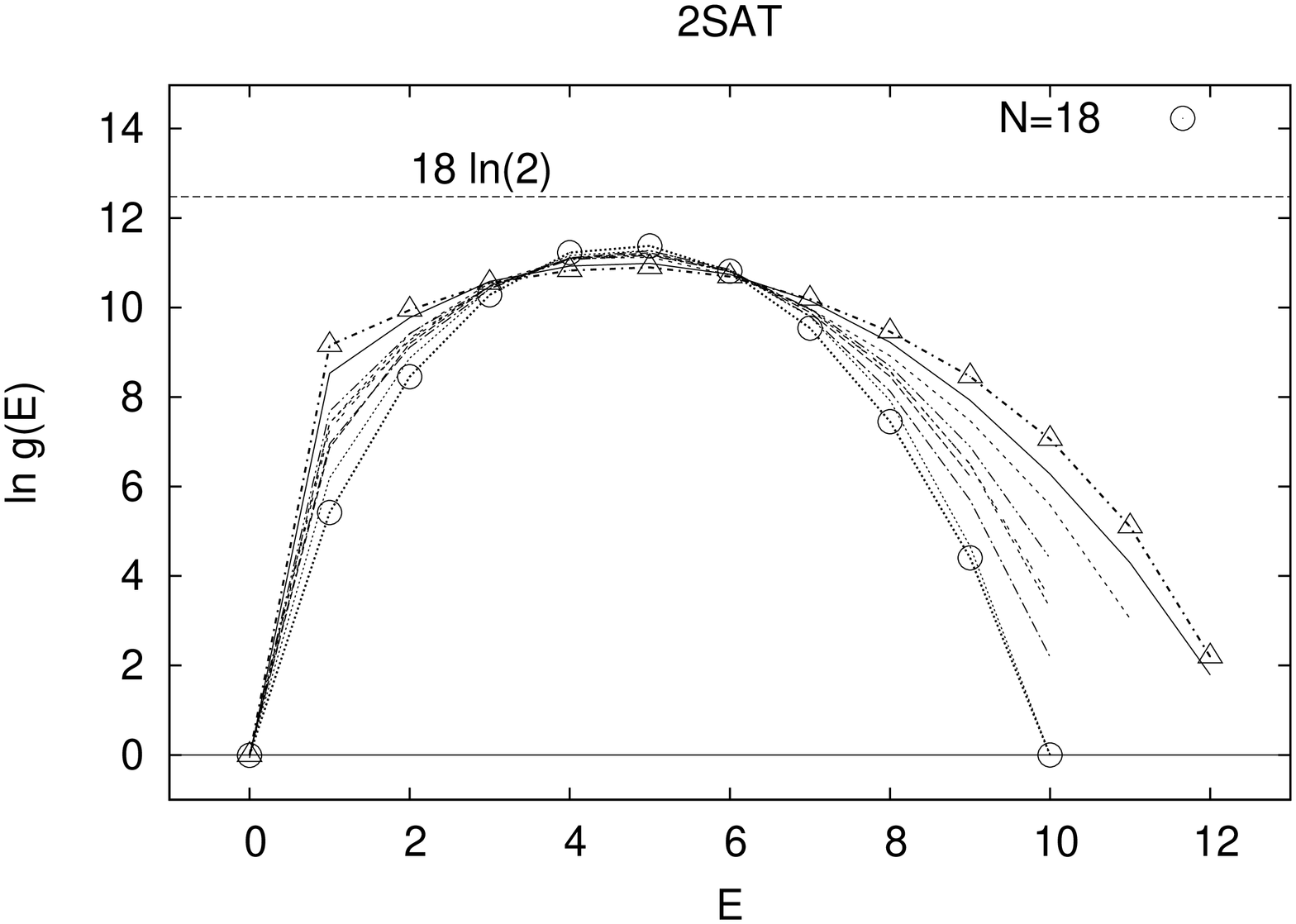} 
  \end{minipage}
  \begin{minipage}[t]{0.5\textwidth}
    \includegraphics[angle=-90,width=8.0cm]{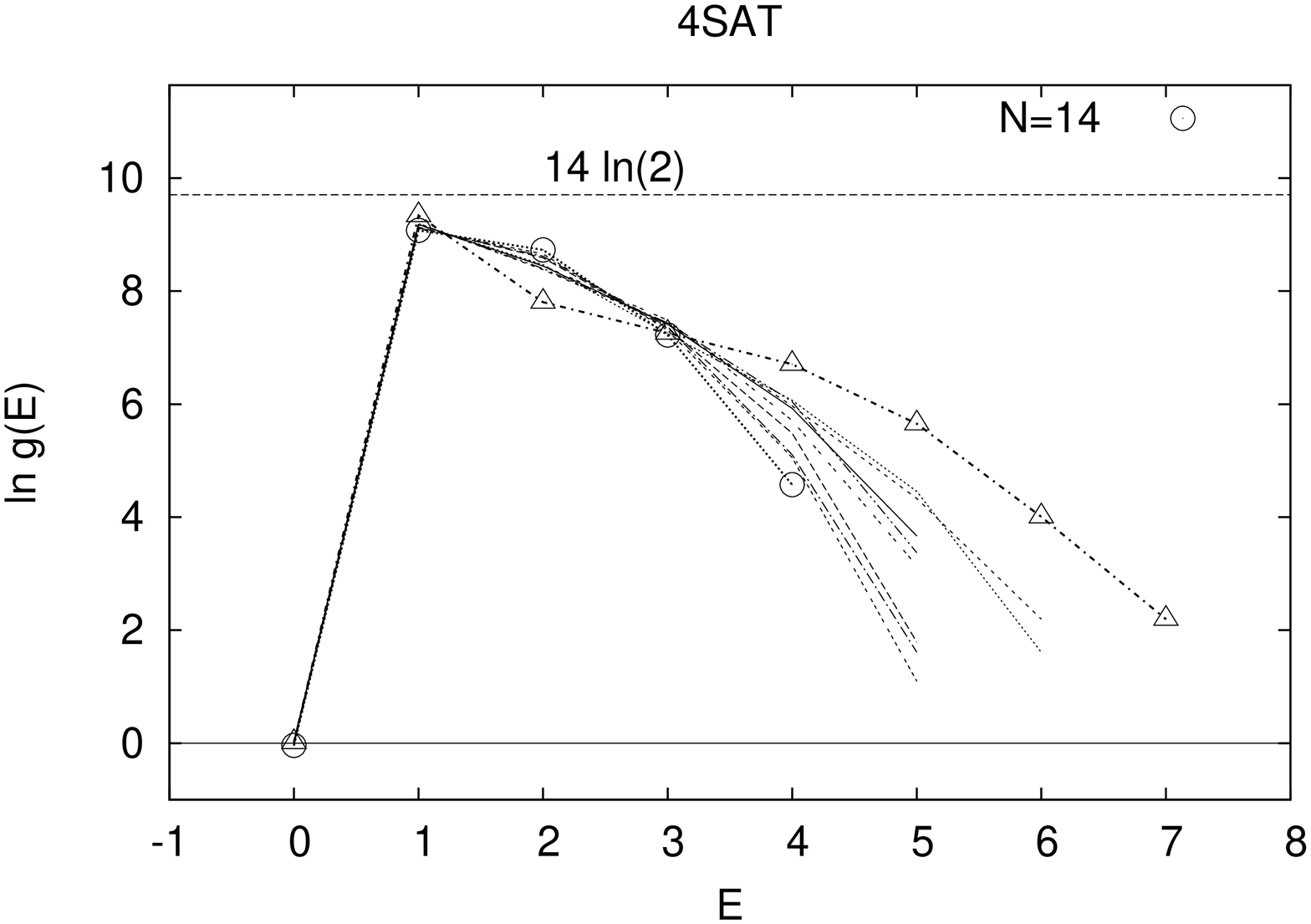} 
  \end{minipage}
\vspace*{0.0cm}\caption{Ten density of states (DOS) 
curves in 2SAT (left) and 4SAT (right) for a  number of spins $N=18$ and $N=14$ respectively.}
\label{fig:dos_a}
\end{figure}
\begin{figure}[t]
  \begin{minipage}[t]{0.5\textwidth}
    \includegraphics[angle=-90,width=8.0cm]{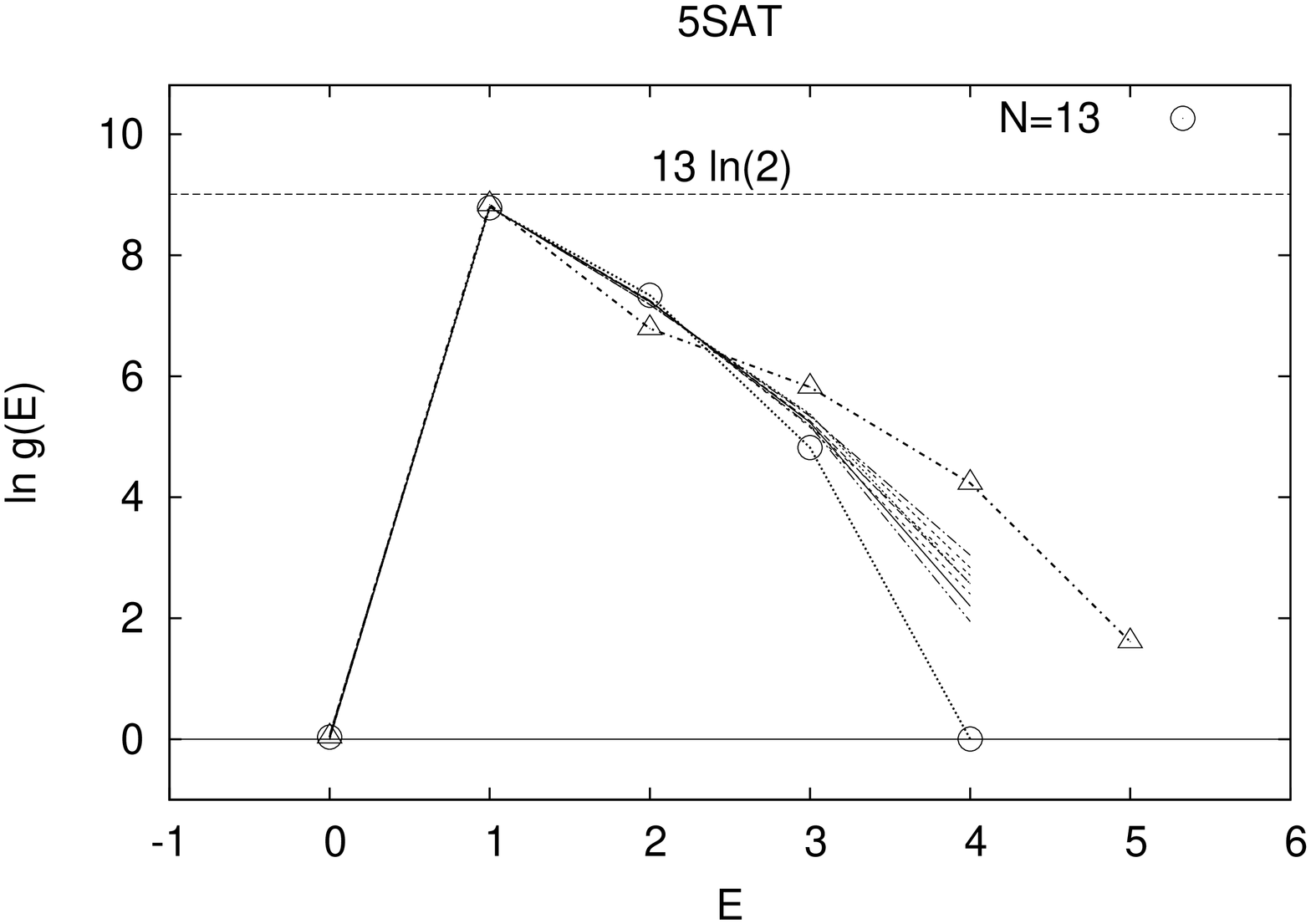} 
  \end{minipage}
  \begin{minipage}[t]{0.5\textwidth}
    \includegraphics[angle=-90,width=8.0cm]{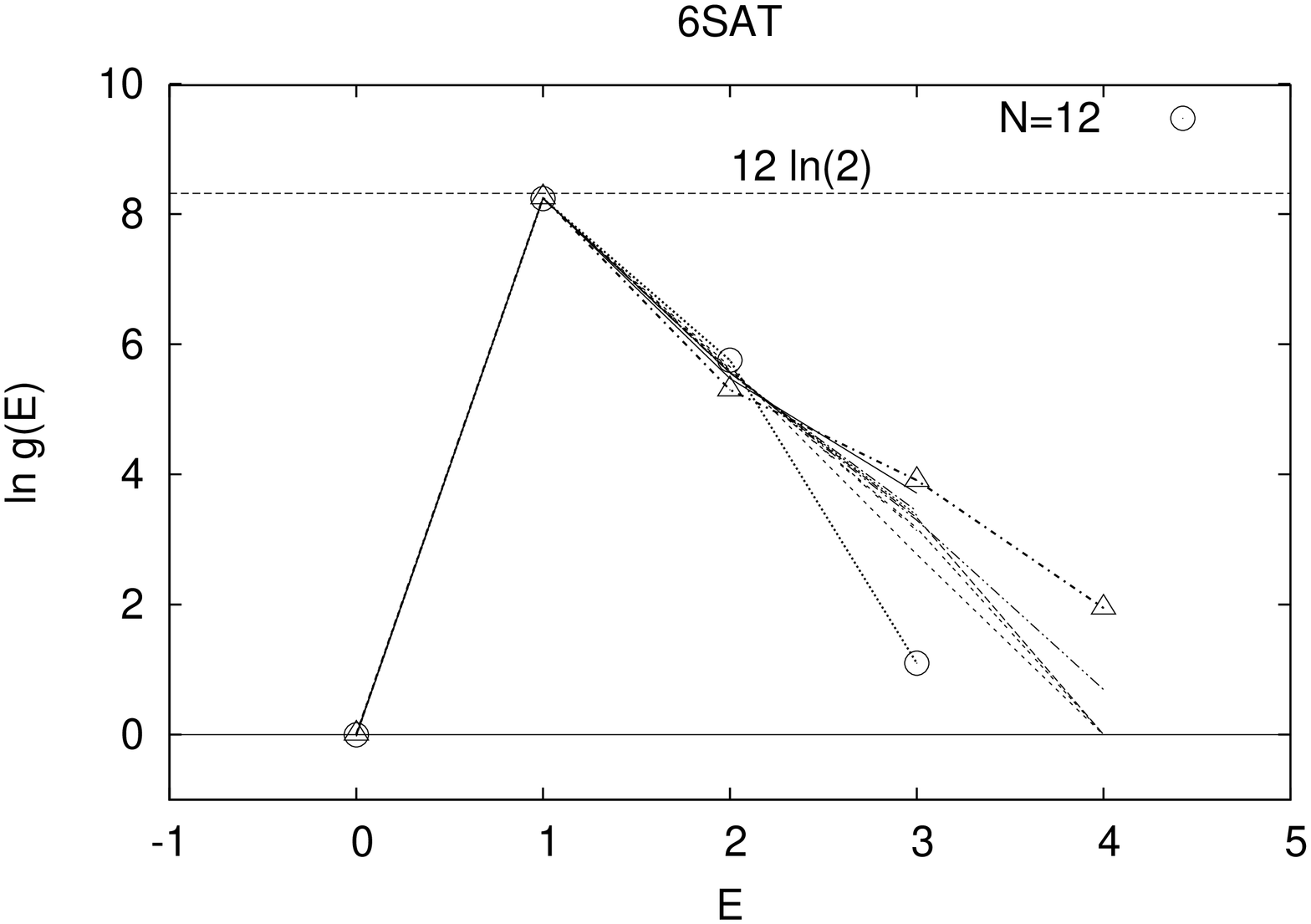} 
  \end{minipage}
\vspace*{0.0cm}\caption{Ten density of states (DOS) 
curves  in 5SAT (left) and 6SAT (right) for $N=13$ and  $N=12$ respectively.}
\label{fig:dos_b}
\end{figure}
\begin{figure}[htb]
\centering
\includegraphics[angle=-90,width=12.0cm]{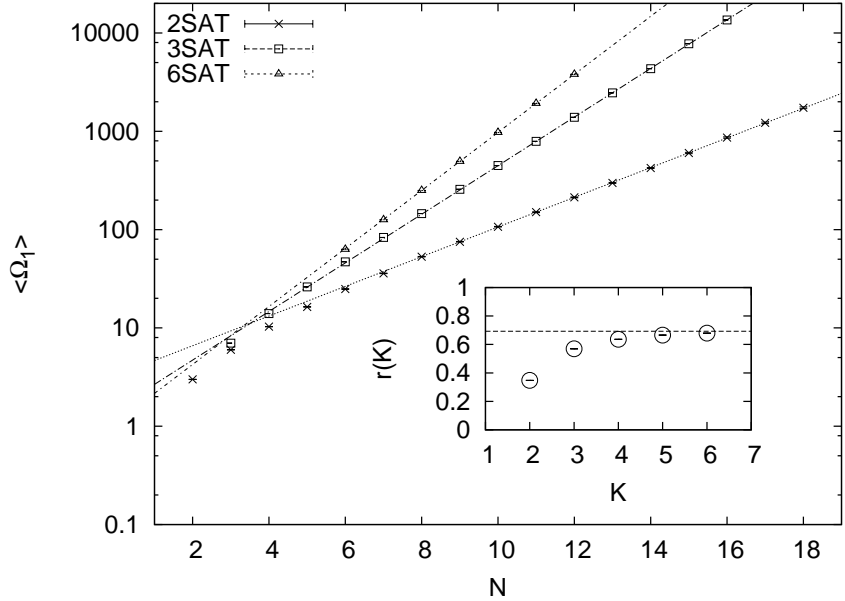} 
\vspace*{0.0cm}\caption{The mean density of states $<\Omega_1>$ on the energy one surface
averaged in the hard problem ensemble $<...>_{\rm HARD}$ for the theories 2SAT, 3SAT
and 6SAT in logarithmic scale as a function $N$.}
\label{fig:omega_one}
\end{figure}
\begin{table}
\centerline{
\begin{tabular}{|c|c|c|}
\hline
 $K$ & $r(K)$ & $\chi^2_{\rm dof}$   \\
\hline
    2 &  0.34762(112) & 0.65    \\
    3 &  0.56918(096) & 1.87    \\
    4 &  0.63620(015) & 0.66    \\
    5 &  0.66574(025) & 2.55    \\
    6 &  0.67934(014) & 0.30    \\
\hline
\end{tabular}}
\label{table1}
\caption{Fit parameters of $\chi^2_{\rm dof}$ fits to $<\Omega_1>$ data with the form 
eq.(\ref{eq:omega_fit}). The rate constants $r(K)$ approach the unstructured search value
${\rm ln}2=0.6931...$ rapidly for large values of $K$.}
\end{table}
For each of the generated problem realizations within the ensemble
$<...>_{\rm HARD}$ and as defined by the partition function $\Gamma(\mu)$
of eq.({\ref{partition_function}}) for $\mu=1$, we calculate the density
of states eq.(\ref{eq:dos}). We determine its mean on the $E=1$ surface
\begin{equation}
<\Omega_1>=<g(E=1)>_{\rm HARD}.
\end{equation}
Selected data for the density of states $g^\eta(E)$
with $\eta=1,...,10$ are displayed in Figures
Fig.(\ref{fig:dos_a}) and Fig.(\ref{fig:dos_b}) for the $K=2,4,5,6$
KSAT theories. They complement the 3SAT data displayed in 
Fig.(\ref{figure_1}). In each case the multiplicity of $E=1$ configurations exhibits a step 
$\Delta {\rm ln}\Omega ={\rm ln}\Omega_1={\rm ln} g(E=1)$ that is of 
magnitude ${\cal O}(N)$ for the given number of spins $N$. Our final numerical 
data for the mean multiplicity of $E=1$ configurations $<\Omega_1>$,
in the theories 2SAT, 3SAT and 6SAT are displayed in Fig.(\ref{fig:omega_one}).
%In all cases and for all values $K$ the numerical data are consistent with an
The numerical data are consistent with an
exponential growth
\begin{equation}
<\Omega_1>~=~ {\rm const}~e^{+r(K)N},
\label{eq:omega_fit}
\end{equation}
for large values of $N$ with finite growth rate constants $r(K)$. 
Subsequently, we performed $\chi^2_{\rm dof}$ fits to the $<\Omega_1>$
data in order to determine the shape of the singularity eq.(\ref{eq:omega_fit})
and to measure values of the rate constants $r(K)$ in KSAT theories with $K=2,3...,6$.
Restricting the fit interval to the cases with $N \ge 10$ we obtain
acceptable $\chi^2_{\rm dof}$-values for the fit. The final rate constants $r(K)$ 
and $\chi^2_{\rm dof}$-values of the fits are contained in Table 1. The $K$-dependence
of the rate constants $r(K)$ is also depicted in the inset of Fig.(\ref{fig:omega_one}).
Starting from a moderate value for the rate constant in 2SAT, $r(K=2)=0.348(2)$, we obtain
$r(K=3)=0.54(1)$ in 3SAT and, beyond $K=3$ the rate constants rapidly approach the unstructured search value
$r={\rm ln}2=0.6931...~$.  For 6SAT the rate constant is $r(K=6)=0.6793(2)$.

\begin{figure}[t]
\centering
\includegraphics[angle=-90,width=12.0cm]{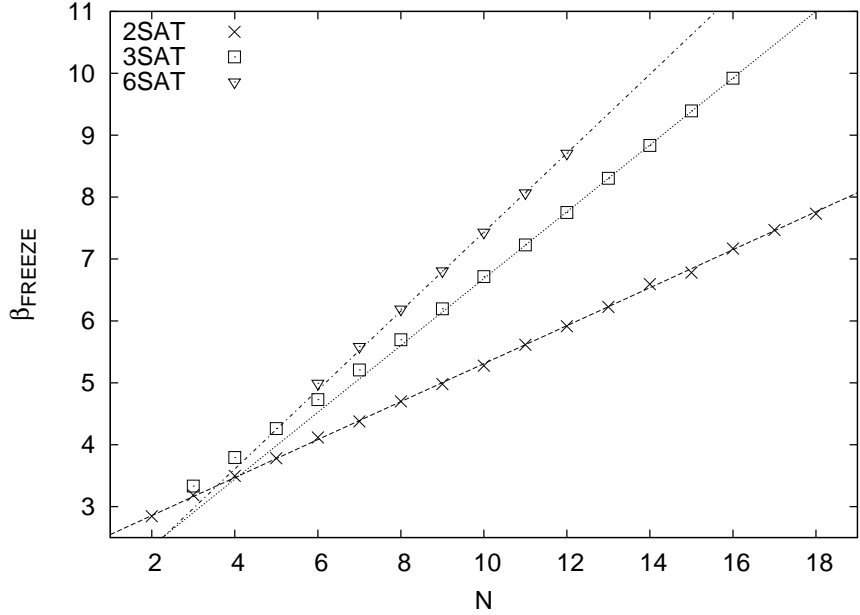} 
\vspace*{0.0cm}\caption{Position of the freezing transition in $\beta_{\rm F}$, the inverse temperature. For
large $N$ values $\beta_{\rm F}$ asymptotically approaches values $\beta_{\rm F}= {\rm ln} \Omega_1$
and thus the slopes the linear behavior approach values $r(K)$ as given in Table 1.}
\label{fig:freezing_temperature}
\end{figure}

A classical statistical model with a density of states $g(E)$, that 
squeezes an exponential large number 
of configurations into the first energy level above the 
ground-state, see the right panel of Fig.(\ref{fig:dos_b}) is certainly a very special 
theory. Let us recall the ferromagnetic Ising model, which in any 
dimension $D$ has a ground-state degeneracy $g(E=0)=2$ as well as a multiplicity $\Omega_1=g(E=2D)=2N$
at the first energy level. Polynomial singularities in $\Omega_1$ are 
the consequence of theories with local interactions. However, 
the class of problems considered here does not possess this property.

\begin{figure}[t]
\centering
\includegraphics[angle=-90,width=10.0cm]{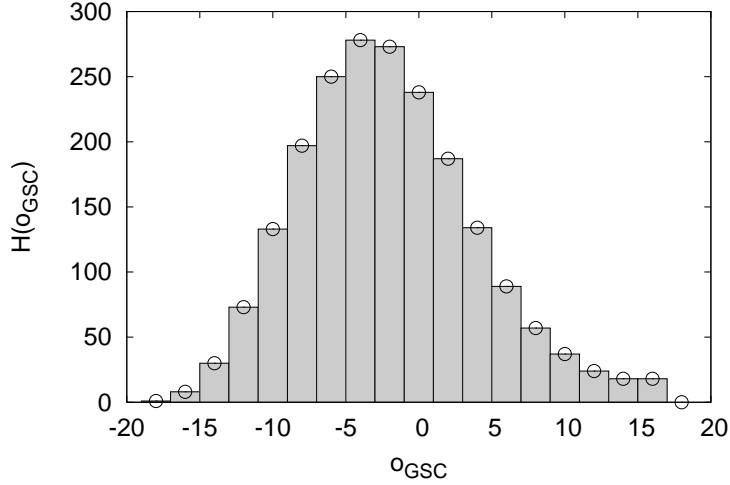} 
\vspace*{0.0cm}\caption{Overlap distribution to the ground state in accord 
with eq.(\ref{eq:olap_formula}) for a single $N=18$ problem in 2SAT.}
\label{fig:olap_2sat_18}
\end{figure}

The spin configurations at energy $E=1$ will have a phase space distribution
and it is interesting to know, whether that distribution is biassed 
towards the ground state configuration. For this purpose we calculate
the overlap to the ground state
\begin{equation}
o_{\rm GSC} = \sum_{i=1}^N s_i s_i^0
\label{eq:olap_formula}
\end{equation}
where $s_i^0$ denotes ground state spins. For purposes of illustration
we display in in Fig.(\ref{fig:olap_2sat_18}) the number histogram $H(o_{\rm GSC})$
for a single $N=18$ problem in 2SAT. We obtain a bell-shaped overlap distribution 
which actually is slightly biassed away form the ground state to the negative half space. 
We note that the histogram carries entries at $o_{\rm GSC}=16$ and thus the 
ground state is accessible via single spin flips from the 
$E=1$ surface. We also have analyzed the connectivity of $E=1$ configurations. Using 
{\it ballistic shooting} we find that any two $E=1$ configurations are connected
by sequences of single spin flips without leaving $E=1$. This is different 
from spin glasses where in general there are several connectivity 
components and corresponding free energy barriers.
Any single spin flip dynamics e.g. Metropolis updates 
can easily explore the $E=1$ surface.
 
\begin{figure}[htb]
\centering
\includegraphics[angle=-90,width=12.0cm]{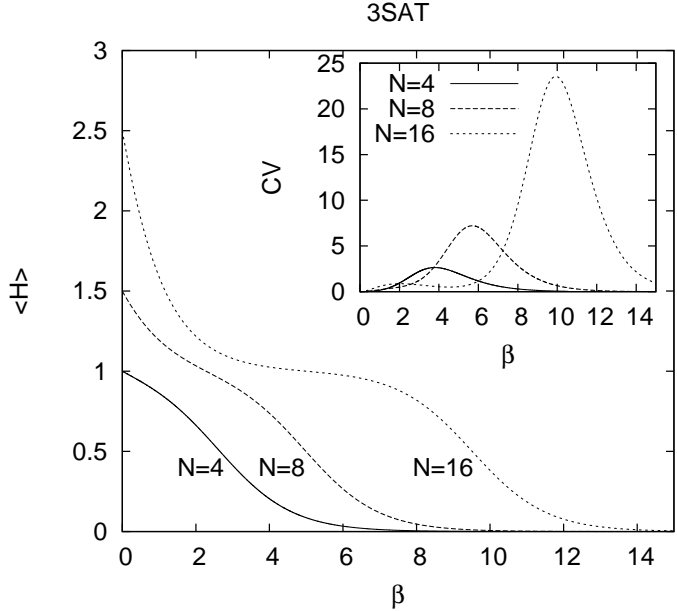} 
\vspace*{0.0cm}\caption{Expectation value $<E>$ of the energy in 3SAT as a function
of the inverse temperature $\beta=1/T$. The inset displays specific heat $<C_V>$ data. At the freezing
transition marked by the position of the maximum of the specific heat an exponentially large number $\Omega_1$
of energy $E=1$ configurations coexists with the single ground-state at $E=0$.}
\label{fig:cost_function}
\end{figure}
We also consider the canonical ensemble eq.(\ref{eq:canonical}).
We calculate the internal energy 
$<E>=<U>_{\rm HARD}$ with $U=\partial_\beta {\rm ln}Z$, as well as the specific heat
$<C_V>(\beta)=<\beta^{2}\partial_\beta U>_{\rm HARD}$, as a function of the inverse temperature
$\beta=T^{-1}$. For 3SAT we display $<E>$ and $<C_V>$ for $N=4,8$ and $N=16$ spins
in Fig.(\ref{fig:cost_function}). A theory with a finite energy gap is expected to possess 
a freezing phase transition at low, possible very low temperatures $T_{\rm F}$ below which and 
for values $T<T_{\rm F}$ the internal energy approaches its asymptotic ground-state 
value $<E>=0$. The numerical data in fact confirm the presence of freezing, with a position as given
by the position of a pronounced peak in the specific heat, see the inset of Fig.(\ref{fig:cost_function}).
Figure (\ref{fig:freezing_temperature}) displays $\beta_{\rm F}=T_{\rm F}^{-1}$ data in 3SAT, which as a function
of $N$ exhibit a blatant linear dependence, see the straight lines in Fig.(\ref{fig:cost_function}). We remark 
that at the freezing point configurations with $E=1$ coexist with a single 
configuration at the ground-state energy. 

\begin{figure}[t]
\centering
\includegraphics[angle=-90,width=12.0cm]{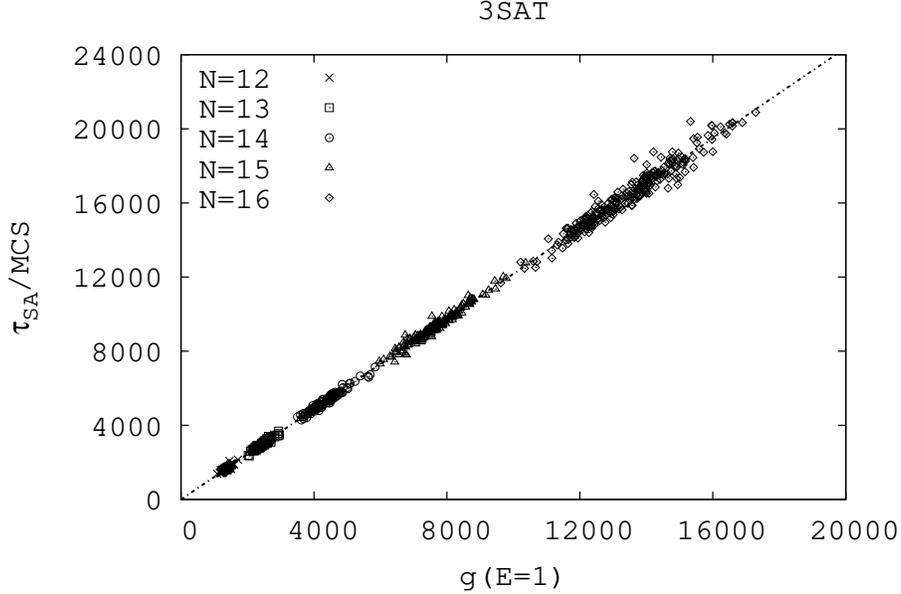} 
\vspace*{0.0cm}\caption{3SAT correlation of run times in simulated annealing eq.(\ref{sa_run_time})
with the density of states $g(E=1)$. We combine run-times at various $N$ into a single plot.
Plotted is a subset of the set of all problems.}
\label{fig:simulated_annealing}
\end{figure}
A popular algorithm within the canonical ensemble for the solution 
of optimization problems is simulated annealing (SA) \cite{simulated_annealing}.
Simulated annealing runs will have to use temperature annealing schedules 
with temperatures low enough to reach the freezing point at 
$T_{\rm F}$ e.g. $T \approx 0.1$ for 16 spins in 3SAT and, then will have to 
explore $\Omega_1$ number of possibilities to finally arrive at the ground-state. 
The process will consume an exponentially large amount of time, if $\Omega_1$ is exponentially
large. We do not expect, that other algorithmic improvements like kinetic 
Monte Carlo methods \cite{Kinetic_Monte_Carlo} can avoid the exponential singularity.

We have implemented simulated annealing for the problem set in 3SAT. 
We use the canonical partition function  eq.(\ref{eq:canonical})
and choose a random initialized spin-configuration. We then perform
local Metropolis spin updates in a multi-spin coded computer 
program \cite{Wansleben_1984,Bhanot_1986}.  
We employ compute time farming on a parallel computer
with a parallel random number generator of 
Marsaglia \cite{Marsaglia_02003}. 
Each annealing trajectory
is started at the very high temperature
\be
T_{\rm 0}~=~100
\ee
and terminates after $1000$ Sweeps i.e., $1000 \times N$ Monte 
Carlo steps where $N$ is the spin number, at the exact temperature
\be
T_{\rm End}=~{1 \over 30}.
\ee 
We use a polynomial temperature schedule
\be
T_{i}~=~ {\rm a}~ i^{-b}
\ee 
where $i$ is the sweep number $i=1,...,1000$ 
and constants $a,b$ are determined
to meet the boundary conditions on the temperature.
For each problem we repeat the annealing trajectories $6400$ times
with different random numbers 
and determine the mean success probability 
$P_{\rm Success}^{SA}$ with  $0 \le P_{\rm Success}^{SA} \le 1 $ of successful 
ground-state searches after the sweep $1000$ has been finished.
Our measure of SA search run-time is  
\be
\tau_{\rm SA}~=~~{{\rm ln}[1-P_{\rm Target}^{SA}] 
\over {\rm ln}[1-P_{\rm Success}^{SA}]}~~\times~~1000~~\times N ~~[{\rm Monte~Carlo~Steps}]
\label{sa_run_time}
\ee
at target success rate one-half : $P_{\rm Target}^{}={1 \over 2}$.
The procedure is repeated for a possible $1000$ 
realizations and at all values of $N$. The correlation of run-times $\tau_{\rm SA}$
with the density of states $g(E=1)$ is linear for 3SAT, as can be inspected
in Fig.(\ref{fig:simulated_annealing}) for a selected set of problems at various $N$.
These run times are quite short. If e.g. at $N=16$ the energy
surface has $16000$ degenerate spin 
configurations a typical number of ${\cal O}(20000)$ Monte Carlo Steps
is sufficient to solve the problem at a success rate of one half. 
However if the target success rate is demanded to be very close to unity
larger times are needed. Our findings imply
that the classical compute time for solving problems with simulated
annealing goes like $\tau_{\rm Classical} \propto e^{+r(K)N}$ with values
$r(K)$ as given in Table 1. It is this kind of singularity a quantum search
has to compete with.

%\newpage

\section{Conclusion}

\begin{figure}[t]
\centering
\includegraphics[angle=-90,width=12.0cm]{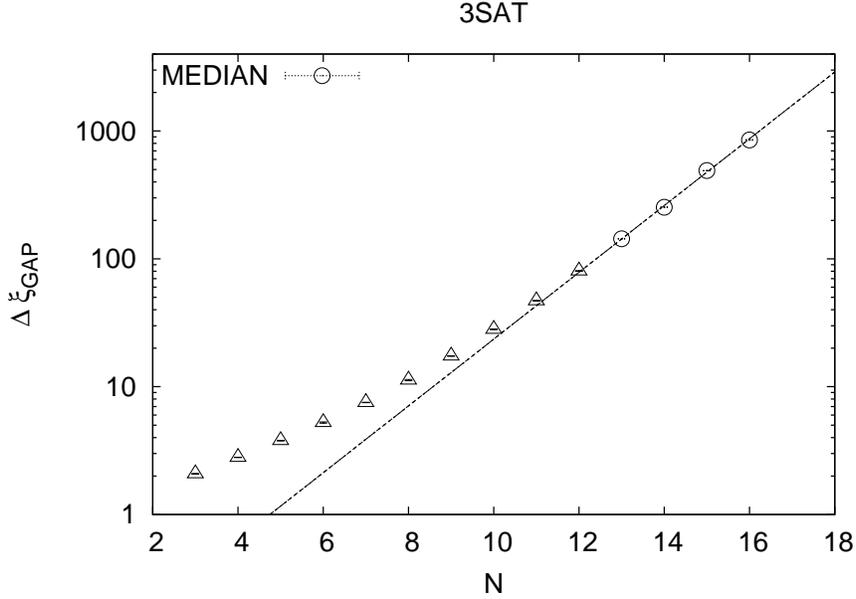} 
\vspace*{0.0cm}\caption{Preliminary quantum gap-correlation length values
$\Delta \xi_{GAP}$ at the quantum phase transition of $Z_{\rm Q}$ in eq.(\ref{quantum_partition}). }
\label{fig:gap_correlation length}
\end{figure}

Within the scope of the present work, we have generated
prototype problem realizations within KSAT 
theories, which under the constraint of a unique
satisfying assignment (USA) at minimal clause number develop
extremal statistical properties. The phase space volume
$\Omega_1$ at the minimal energy gap is exponentially large and 
likewise for a given KSAT theory maximal. The idea was
formulated in $2005$ by Znidaric \cite{Znidaric_2005} but in absence
of efficient Monte Carlo methods it was not worked out at minimal clause 
number and at large values of the rate constants $r(K)$. 
The class of problems as presented here exemplifies
our current understanding of physical search complexity in random systems
in a straight and simple way: A single ground state 
is hidden in an exponentially large phase space volume at the first energy gap.
For the theories with large $K$ almost all spin configurations are
collapsed to the $E=1$ surface, except the one ground state configuration
at $E=0$. In this situation there exists no distance measure or cluster
property which within the $E=1$ surface would allow the detection
of a direction, as to where the ground state could be searched 
for. Representatives of the ensemble $<...>_{\rm HARD}$ can be 
obtained at request from the author.

The given problems at $K=3,4,5$ and $K=6$ in this work
are constructed on problem Hamiltonians that
contain higher order interactions of spins 
like $a_{i,j,k}s_i s_j s_k$. From a physics point
of view it would possibly be nicer to eliminate such unphysical
couplings and stay with only 2-point spin couplings, as well as 
magnetic fields. We mention that all the Hamiltonians at $K \ge 3$
can be transformed via polynomial transformations
to Maximal Independent Set (MIS),see \cite{mis}, which in fact 
can be represented by 2-point and magnetic field spin couplings only. 
It is plausible to assume that these after polynomial transformation retain
their ``hardness''.

The design of problem realizations with specific 
properties facilitates the subsequent study of proper defined 
search efficiency's in processes, that can possibly be 
implemented on a physical device e.g., a quantum computer. For purposes 
of illustration we mention here quantum annealing within the 
quantum partition function $Z_{\rm Q}$ of eq.(\ref{quantum_partition}). 
Search times for ground-state 
calculations via quantum annealing are expected to be 
bounded by below through a gap-correlation length
$\Delta \xi_{GAP}$, which is determined from spin-spin correlations
along the imaginary Trotter Suzuki time of $Z_{\rm Q}$ at the quantum 
critical point. For 3SAT we present in Fig.(\ref{fig:gap_correlation length}) 
preliminary numerical results for $\Delta \xi_{GAP}$ in the median 
average of the hard problem ensemble. 
The data, as indicated by the straight line in the figure, show in fact
also an exponential singularity $\Delta \xi_{GAP}\propto {\rm exp}[+r_QN]$ 
of a similar type as in eq.(\ref{eq:omega_fit}), that now is governed 
by a quantum rate constant $r_Q \approx 0.60(1)$, a value that is close
to $r(K=3)=0.569(1)$ of Table 1. The caveat however is, that in presence
of a Landau Zener avoided level crossings quantum run-times 
for linear quantum annealing schedules 
show a quadratic singular behavior 
$\tau_{\rm Quantum} \propto \Delta \xi_{GAP} ^2$ 
\cite{Zener_1932}, which leaves the quantum search 
efficiency far behind the classical
search. Similar exponential singularities at smaller
values of $r_Q$ were already observed for quantum 3SAT on a set of 'weaker' 
problems \cite{neuhaus_02011}. A detailed study of quantum search complexities
on the set of hard problems in 2SAT has just been 
completed \cite{neuhaus_2sat_02014} and complements the less 
physical findings of this work.  

Finally we mention that the spin numbers $N$ in this work are
embarrassing small, as the Monte Carlo search on the problem set
consumes exponentially large resources. We can safely say that with
current methods it is not possible to generate a corresponding
ensemble of problems even for spin numbers as small as $N=30$.
It is however not excluded, that single problem representatives 
can be found by clever heuristic construction. We emphasize
that we do not want to give up the ensemble property because otherwise
we would be studying arbitrary mathematical problems. This will be relevant
for search complexity distributions which are expected to
exhibit ensemble properties.

{\bf Acknowledgement:} Calculations were performed under the 
VSR grant JJSC02 and an Institute account SLQIP00 at J\"ulich
Supercomputing Center on various computers.

%\newpage

\end{document}